\documentclass{article}

\usepackage{arxiv}

\usepackage[utf8]{inputenc} 
\usepackage[T1]{fontenc}    
\usepackage{hyperref}       
\usepackage{url}            
\usepackage{booktabs}       
\usepackage{amsfonts}       
\usepackage{nicefrac}       
\usepackage{microtype}      
\usepackage{cleveref}       
\usepackage{lipsum}         
\usepackage{graphicx}
\usepackage{doi}
\usepackage{xcolor}
\usepackage[numbers]{natbib}

\newbox{\orcid}\sbox{\orcid}{\includegraphics[scale=0.06]{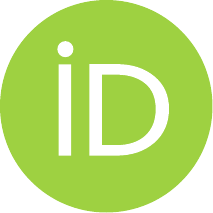}} 

\newcommand{\onkar}[1]{\textcolor{teal}{#1}}

\title{From streaks to synergies: A multi-scale analysis of performance and scoring in the NBA}

\usepackage{authblk}

\author[1]{\href{https://orcid.org/0009-0005-4339-9925}{\usebox{\orcid}\hspace{1mm}Malvina Bozhidarova}}
\author[2]{\href{https://orcid.org/0000-0003-0798-9678}{\usebox{\orcid}\hspace{1mm}Yanpei Cai}}
\author[3]{\href{https://orcid.org/0009-0001-6605-7603}{\usebox{\orcid}\hspace{1mm}Ricardo M.S. Carvalho}}
\author[4,5]{\href{https://orcid.org/0009-0001-6727-3316}{\usebox{\orcid}\hspace{1mm}Daniele Cirulli}}
\author[6]{\href{https://orcid.org/0009-0002-6342-759X}{\usebox{\orcid}\hspace{1mm}Quentin Dehaene}}
\author[7]{\href{https://orcid.org/0009-0009-3682-4814}{\usebox{\orcid}\hspace{1mm}Martin Diaz}}
\author[8]{\href{https://orcid.org/0009-0004-6797-726X}{\usebox{\orcid}\hspace{1mm}Alexandra Krasnokutskaya}}
\author[9]{\href{https://orcid.org/0009-0001-1172-6865}{\usebox{\orcid}\hspace{1mm}Bernardo Pereira}}
\author[10]{\href{https://orcid.org/0000-0002-5177-0836}{\usebox{\orcid}\hspace{1mm}Onkar Sadekar\thanks{Corresponding author: \href{mailto:sadekaronkar@gmail.com}{sadekaronkar@gmail.com}}}}
\author[11,12]{\href{https://orcid.org/0000-0001-9646-6232}{\usebox{\orcid}\hspace{1mm}Federico Battiston\thanks{Corresponding author: \href{mailto:battistonf@ceu.edu}{battistonf@ceu.edu}}}}

\affil[1]{Institute of Mathematics and Informatics, Bulgarian Academy of Sciences} 
\affil[2]{Artificial Intelligence and its Applications Institute, School of Informatics, The University of Edinburgh, Edinburgh, United Kingdom} 
\affil[3]{LASIGE, Faculty of Sciences, University of Lisbon, Lisbon, Portugal} 
\affil[4]{Physics Department and INFN, University of Rome ``Tor Vergata'', Rome, Italy} 
\affil[5]{``Enrico Fermi'' Research Center, Rome, Italy} 
\affil[6]{New York University, New York, USA} 
\affil[7]{Department of Condensed Matter Physics, University of Barcelona, Barcelona, Spain} 
\affil[8]{Complexity Science Hub, Vienna, Austria}
\affil[9]{Department of Mathematical Sciences, Politecnico di Torino, Torino, Italy} 
\affil[10]{Human Evolutionary Ecology Group, Department of Evolutionary Anthropology, University of Zurich, Zurich, Switzerland} 
\affil[11]{Department of Network and Data Science, Central European University Vienna, Vienna, Austria} 
\affil[12]{Department of AI, Data and Decision Sciences, Luiss University of Rome, Rome, Italy} 

\begin{document}
\maketitle

\begin{abstract}
Modern play-by-play data make it possible to test long-standing intuitions about basketball with the same statistical rigour now routinely applied to other professional sports. Using play-by-play data from 7,054 regular-season and 504 playoff NBA games spanning the 2019/20–2024/25 seasons, we provide quantitative insights into scoring patterns and the performance of individual players and teams through methods from statistics, network science, and complexity science. Our findings offer an evidence-based perspective on in-season and in-game performance that can inform coaching strategies, player evaluation, and tactical decision-making.
\end{abstract}


\section{Introduction}

Recent advances in digital tracking and data acquisition have transformed professional sport into a useful laboratory for the quantitative study of human performance and collective behaviour~\cite{mariani2024collective}. Two decades ago, the Moneyball revolution in professional baseball demonstrated that statistical reasoning could overturn traditional managerial intuitions and reshape the logic of player evaluation~\cite{lewis2004}. Since then, individual sports such as tennis~\cite{radicchi2011}, as well as team sports such as football~\cite{buldu2019defining, carstens2026evolution}, American football~\cite{romer2002s}, and cricket~\cite{sadekar2024individual}, have similarly embraced analytics to refine individual and team strategies and in-game decisions, often in tension with the accumulated wisdom of coaching traditions~\cite{neuhaus2024}. The availability of large archival databases and the development of analytical tools has extended this approach well beyond traditional sport arenas, enabling systematic training and the discovery of new strategies in intellectual games, from videogames~\cite{herbrich2006trueskill, avontuur2013player} to Go~\cite{silver2016mastering} and chess~\cite{vaci2017, chowdhary2023, de2023quantifying, barthelemy2025fragility}.

Basketball is one of the most prominent sports in the world, captivating millions of fans across continents with its fast-paced action, global reach, and rich competitive history. The National Basketball Association (NBA) is widely regarded as the premier professional basketball league in the world, showcasing the sport's highest level of talent, competition, and global influence. Basketball strategy and decision-making have historically been informed by a blend of empirical evidence and conventional wisdom, from Phil Jackson's 40--20 rule, which suggests that only teams achieving 40 wins before 20 losses are credible championship contenders, to enduring beliefs in defensive dominance, momentum, and the ``hot hand''. The availability of large-scale match data has made it possible to scrutinise these intuitions systematically~\cite{terner2021}. Research has formalised possession-based efficiency metrics as a common evaluation metric~\cite{kubatko2007}, tested and ultimately partially rehabilitated the hot hand effect after correcting for a streak selection bias in the original analysis~\cite{gilovich1985,bareli2006,miller2018}, exploited optical tracking data to model possession value and spatial shooting structure~\cite{cervone2016}, and characterised team strategy through ball-movement networks~\cite{fewell2012,sampaio2015}. Despite this breadth, the sequential structure of team-level outcomes across full seasons and games, the degree to which winning is genuinely self-reinforcing, and which team features matter to achieve such results, remain underexplored~\cite{terner2021}. In this work, we provide a data-driven lens to some of these long-standing intuitions, tracking patterns of individual and team success across NBA seasons, using statistical, network and complexity science approaches to examine the conditions under which teams win matches and ultimately the championship.

In this work, we leverage six seasons of NBA play-by-play data (2019/20--2024/25) to map persistence and momentum at three nested scales. At the season scale, we test whether statistically significant signatures beyond randomized null models exist for the presence of winning streaks, home advantage, travel and rest, and we contrast win-rate-based and network-based summaries of team strength and playoffs performance. At the game scale, we investigate team scoring runs, quantify the timing of lead changes and the score-margin thresholds at which the outcome becomes effectively determined, and characterise the temporal profile of individual scoring streaks. Finally, we examine
NBA line-ups, identifying synergistic and anti-synergistic combinations of players.

\section{Data}
\label{sec:data}

We used NBA data collected in Ref.~\cite{data_source} for our analysis. The data consists of $7054$ matches in the regular season and $504$ matches in the playoffs across 6 seasons from $2019/20$ to $2024/25$. We use fine-grained play-by-play statistics such as team line-ups, team-scores, points scored by individual players, for around $3.6$ million timestamped events allowing us to identify substitution and shot made events. By aggregating data at different scales (quarter, game, or season level), we get individual and team performance metrics.


\section{In-Season analysis}
\label{sec:season}

The first part of this report focuses on season-level patterns of team performance, examining temporal dynamics of wins, the effects of home advantage, travel and rest, and strength of schedule, and how these factors relate to overall success across an NBA season.

\subsection{Team streaks across season}

Hot streaks -- periods of unusually sustained success or high performance -- have long fascinated researchers across domains, ranging from science to sports~\cite{liu2018hot, miller2018, janosov2020success, ram2022significant}. In the media discourse about NBA, winning streaks are often viewed as indicators of a team’s competitive form, reflecting periods during which tactical cohesion, player performance, and collective confidence align to produce sustained success. Understanding whether such streaks arise from genuine changes in performance or simply reflect random fluctuations remains a central challenge in the study of human behavior and complex systems.

In sport, success tends to compound itself. A remark by John Wooden, who led UCLA basketball to ten national championships in twelve years, captures this intuition precisely: ``Winning breeds winning''. Across the seasons analyzed here, this reinforcement pattern is born out quantitatively: the conditional probability of winning a game immediately following a victory is $P(W|W) = 54.5\%$, a value above the $50\%$ expected under any memoryless model and indicative of a genuine autocorrelation in team performance.

\begin{figure}[!h]
    \centering
    \includegraphics[width=\linewidth]{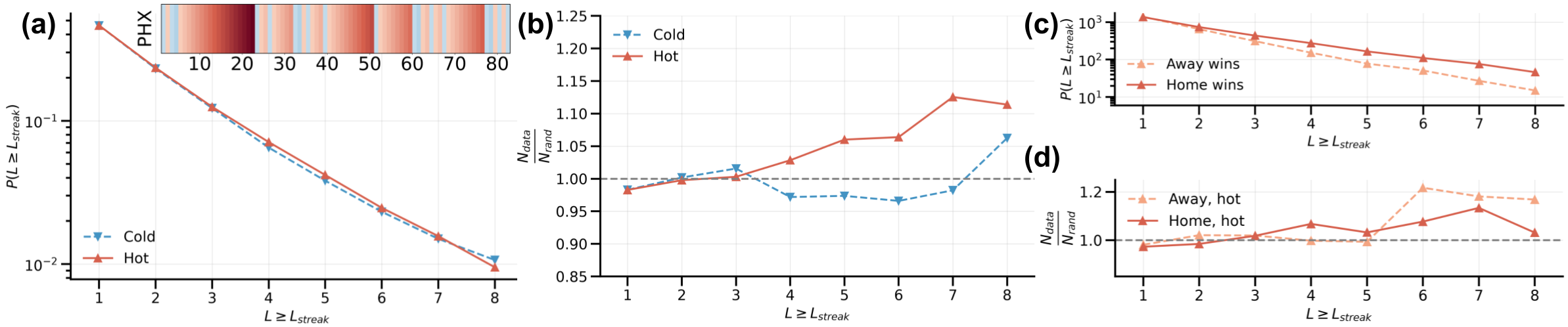}
    \caption{\textbf{Team streaks across season.} \textbf{(a)} Probability distribution function $P(L \geq L_{streak})$ denoting hot (red thick line) and cold (blue dotted line) streaks of $L$ or more consecutive wins or losses respectively across a season. Note that the y-axis is in a logarithmic scale. The inset illustrates a season with a long hot streak (Phoenix Suns PHX, 18 consecutive wins in 2021/22). The shade gets redder or bluer for each consecutive win or loss. \textbf{(b)} Ratio of the number of observed streaks of length $N_{data}$ to the number of expected streak length $N_{rand}$ computed by a random shuffle of game outcomes for each team in a given year. Values above 1 indicate that long streaks occur more often than expected by chance. \textbf{(c)-(d)} Same trends as panels (a) and (b), with hot streaks computed separately for home games only (dark red), and away games only (light red).}
    \label{fig:within_season_hot_streaks}
\end{figure}

To capture the presence of temporal correlations, in Fig.~\ref{fig:within_season_hot_streaks} (a) we show the distribution of winning and losing streak lengths aggregated across all NBA teams and seasons. While most streaks are short (mean $\mu = 2.17$, standard deviation $\sigma = 1.78$), notable extremes exist: 18 consecutive wins by the Phoenix Suns in $2021/22$ (shown in inset of Fig.~\ref{fig:within_season_hot_streaks} (a)) and 28 straight losses by the Detroit Pistons in the season of 2023/24. Long winning or losing streaks, however, can arise purely by chance even in a memoryless system.
Vince Lombardi, a famous NFL coach who coached Green Bay Packers to multiple NFL championships and Super Bowl wins commented that ``Winning is a habit. Unfortunately, so is losing.'' hinting that both wins and losses cluster together. In Fig.~\ref{fig:within_season_hot_streaks} (b) we compare observed streak frequencies against a null model to find that even though wins cluster together, the same pattern does not hold for consecutive losses. For short winning and losing streaks ($L \leq 4$) the ratio remains close to 1, signifying no deviation from a random expectation. Instead for longer streaks ($L \geq 5$), the ratio for wins rises steadily, reaching $\approx 1.15$ for $L = 6$, confirming that long winning streaks occur more often than chance would predict. However, the ratio for longer streaks of losing is below or close to 1 for $L \leq 7$ compared to the randomized pattern and only crosses $1$ to reach $1.05$ for $L=8$, indicating lack of statistical significance for clusters of losses. 
In many sports, home advantage plays a crucial role in performance owing to familiarity with the conditions as well as the support of local fans. Some NBA arenas are particularly famous for the intensity of the support by their fans, such as the \textit{AT\&T Center Fortress} of the San Antonio Spurs, who had an impressive 40 wins against 1 loss record at home in 2015/16 season, and a modest 27 wins against 14 losses in away locations. Overall, in the seasons considered in our dataset, teams win at a higher rate (0.55) at home, compared to winning at away locations (0.45). Building on this idea, in Fig.~\ref{fig:within_season_hot_streaks} (c), we look at the patterns of consecutive wins while considering only home games and only away games. Teams tend to have a larger number of long winning streaks at home, compared to long winning streaks at away locations. However, when we account for expected number of long winning streaks based on random chance, we observe that the difference between home and away conditions ceases. In both cases, the ratio rises steadily for longer streaks ($L > 5)$ suggesting that the reinforcement of winning does not differ in home contexts compared to away games.

To summarize, at the aggregate level, NBA teams appear to display a mild but significant tendency to win games in clusters of 6 consecutive games or more, but the losses are typically not clustered. Surprisingly, long winning streaks are equally common in both home and away contexts, once win rate at home and away is taken into account.

\subsection{Home advantage and impact of fatigue on team performance}

\begin{figure}[!h]
    \centering
    \includegraphics[width=\linewidth]{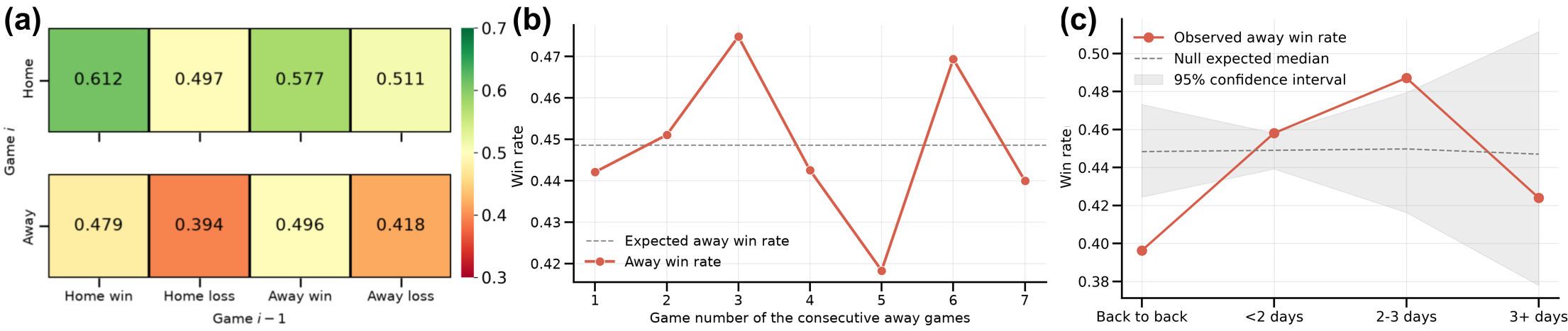}
    \caption{\textbf{Home-court advantage, travel and rest impact on win rate.} 
    \textbf{(a)} Conditional win rate for game $i$ given the location and outcome of game $i-1$, separated by whether game $i$ is played at home (top) or away (bottom). 
    \textbf{(b)} Win rate for the $k^{th}$ consecutive away game also popularly known as road trip games, where $1\leq k \leq 7$  relative to the overall away win fraction (dashed line). 
    \textbf{(c)} Win rate by rest time before a game, relative to a null-model expected median and its $95\%$ confidence interval (shaded band) derived from random shuffling of game outcomes.}
    \label{fig:within_season_away_wins}
\end{figure}

Several factors are deemed important for a team to enter a streak. To further characterize the presence of memory in the system, in Fig.~\ref{fig:within_season_away_wins} (a) we show the win rate for home (top) and away (bottom) games as a function of the previous game.  Results demonstrate both strong home-court advantages and outcome dependency. As mentioned, winning drives momentum, raising the next game's win rate by 11.5\% for consecutive home games and 7.8\% for consecutive away games. However, winning is not the only driver, as travel and rest impact performances. Teams remaining at home after a win show a 4.5\% performance edge over those returning from a road game. Conversely, road trips (i.e. consecutive away games) are particularly challenging. Indeed, returning home immediately after an away loss leads to a 9.3\% improvement in win rate compared to playing another away game.

 In sports media surrounding the NBA, a phenomenon called road trip fatigue is often acknowledged: the more consecutive away games a team plays, the more likely it is to loose these games. In Fig.~\ref{fig:within_season_away_wins} (b) we show the win-rate in consecutive away matches. It contradicts this expected pattern as it shows little diversion from the expected win rate of an away game, independently of its order in the road trip (the same trend is observed if we limit the analysis only to long road trips, e.g. at least 5 away games). While, the third consecutive away game appears to be more frequently won than the first or the second, only the fifth consecutive road game has a notably lower win rate. Although, such games (road trip games number 5 or above) represent only 6.4 \% of all away games).
 
Finally, rest matters. In Fig. \ref{fig:within_season_away_wins} (c) we plot the win-rate as a function of the time from the previous match. It shows that back-to-back games (i.e., consecutive games in consecutive days) show a significantly lower win rate than expected independently of rest time (44\% vs the expected 50\%). This effect is likely emphasized by the modern tendency to protect star players from injury and fatigue by resting them on the second match of a back-to-back, weakening the team for that game.
Such results explain why the NBA seeks to balance out the number of back-to-back games between teams and across seasons, as an increased number is associated with a team's final record. Interestingly, after three or more days of rest, the win rate decreases slightly. Although this trend lies within the confidence intervals and may therefore not be statistically significant, it is consistent with the commonly cited notion of teams becoming rusty after an extended break.

\subsection{``40 before 20'', regular season trends and playoffs performance}

\begin{figure}[!h]
    \centering \includegraphics[width=\linewidth]{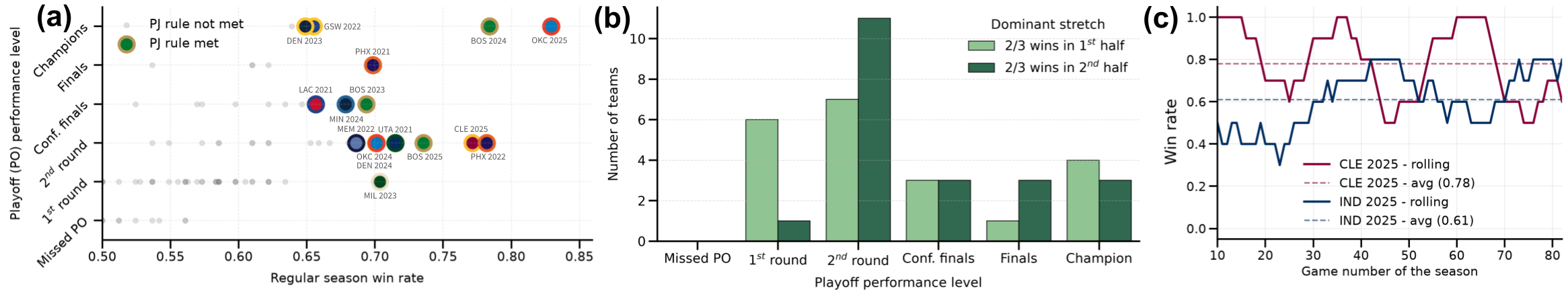}
    \caption{\textbf{Regular season performance and within-season streak patterns relative to playoff outcomes.}
    \textbf{(a)} 
    Playoff performance level by regular-season win rate, highlighting team-seasons that met the "Phil Jackson (PJ) rule" (a team should reach 40 wins before 20 losses in order to make it to playoffs), with each contender marker coloured by team. 
    \textbf{(b)} 
    Number of teams reaching each playoff performance level, split by whether their dominant 2/3-win stretch occurred in the first or second half of the regular season. 
    \textbf{(c)} 
    Rolling win rate (10-game window) across the season for two contrasting examples (Cleveland Cavaliers and Indiana Pacers in 2024-2025 season), shown against each team's overall season average (dashed lines).
    }
    \label{fig:playoffs}
\end{figure}

The NBA is a long tournament divided into 2 conferences which are further divided into 6 divisions. Each team plays 82 games and faces every other team at least twice and maximum 4 times. Phil Jackson, one of the greatest coaches in the history of NBA, famously made the observation that in order to be considered a serious contender for the NBA championship, a team needs to win 40 matches, before losing 20 matches. Indeed in the last 47 seasons (1980-2026), only 5 teams managed to win the championship despite not following the rule. 
Fig. \ref{fig:playoffs} (a) illustrates this dynamic by showing the playoff performance of teams with at least a 50\% win rate over the five seasons at hand. Notably, four out of the five champions met the Phil Jackson rule; however, many teams that met the criteria still failed to make deep playoff runs. While the rule is compelling, only half of the qualifying teams reached the Conference Finals, the expected performance of a legitimate contender. This suggests the rule may be insufficient on its own. Crucially, it only considers the beginning of the season (the first 60 games). Naturally, our interest lies in seeing whether the second half of the season is more vital for building momentum, mirroring the game-to-game dynamics observed previously.

Fig. \ref{fig:playoffs} (b) displays the postseason performance of teams that showed a dominant stretch (winning at least two-thirds of their games in either the first or second half) based on when that stretch occurred. Within this five-year sample, every team that won two-thirds of their games (the same rate of the 40-20 rule) in either half successfully made the playoffs. We also observe no unsuccesful late-season comebacks or catastrophic collapses. Additionally, 95\% of the teams with a dominant second half reached at least the second round, while only 71\% of the teams that had a dominant first half did so, suggesting that late-season momentum protects a team from first failure. However, this logic does not extend indefinitely. When looking at teams reaching the Conference Finals or beyond, 38 \% of teams with a dominant first half qualified, compared to 42\% of teams with a dominant second half. Ultimately, this margin is too narrow to definitively declare late-season momentum a superior predictor of championship success.

Because of such variance during the long NBA season overlooked in team's raw record matters, we decided to look at momentum and overall progress during the season. Fig. \ref{fig:playoffs} (c) illustrates this by tracking the 10-game rolling win rate of two teams with contrasting trajectories during the same season. In 2024-2025 the Cleveland Cavaliers put together a dominant regular season with an overall 78\% win rate, driven by an impressive 85\% win rate in the first half of the schedule. However, their record dropped to 73\% in the second half: still an elite performance, but a significant decline. Conversely, during the same season the Indiana Pacers started slowly but picked up the pace as the year progressed. After averaging a 53\% win rate in the first half, the Pacers improved their win rate by 15\% ( to 68\%) in the second half, while the Cavaliers decreased theirs by 15\%. When these two teams met in the Eastern Conference Semifinals, the ascending Indiana Pacers upset the Cavaliers. Examples like this reinforce the idea that late-season momentum carries immense weight, proving that high baseline performance alone is not always enough. Under such differences, it might be informative to consider not only the number of wins, but also the strength of the opponents that they won against, as well as the team momentum, specially during the last part of the regular season heading into the playoffs. A network-based approach can help us characterize the true strength of a team beyond just the number of wins.

\begin{figure}[!h]
    \centering \includegraphics[width=1\linewidth]{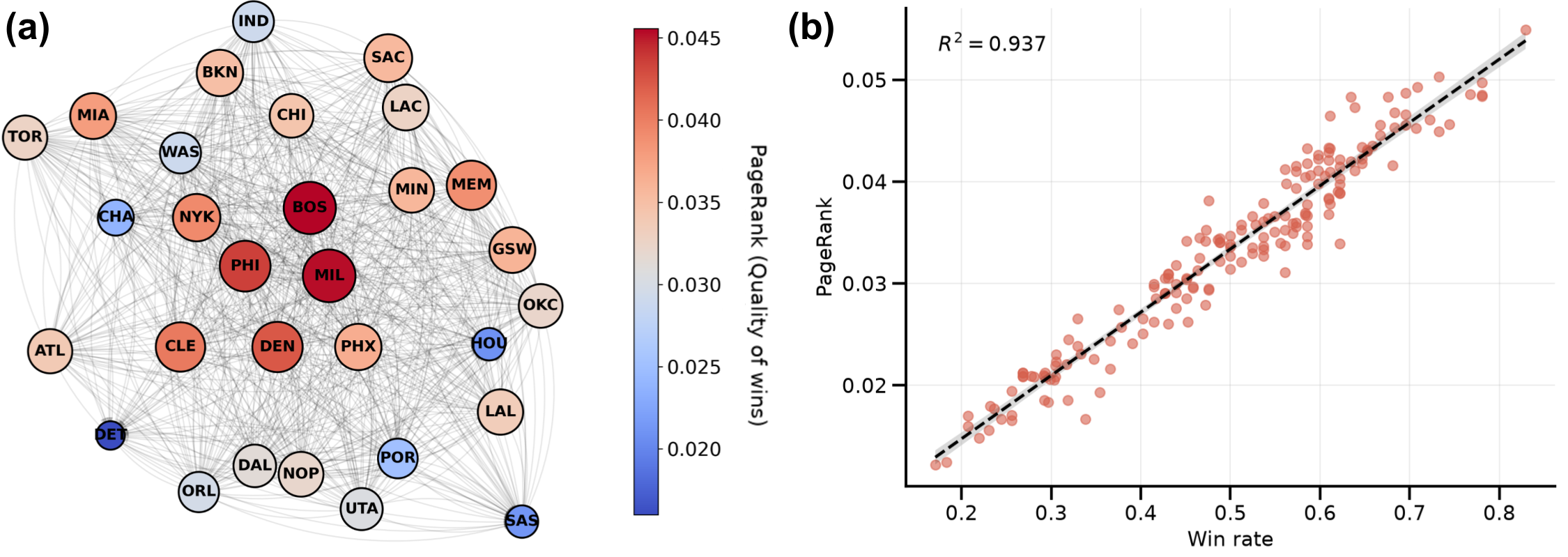}
\caption{\textbf{Network-based approaches for ranking and team performance}
\textbf{(a)} Directed weighted network for the 2022 season, where a directed edge from team $i$ to team $j$ indicates a win by $i$ over $j$, with weight equal to the number of such wins. Node size encodes win rate; node colour encodes PageRank centrality (red: high, blue: low). \textbf{(b)} PageRank centrality versus standard win rate across all team-seasons. The dashed line is a linear regression fit; shaded band shows the 95\% confidence interval.}
\label{fig:network}
\end{figure}

In Fig.~\ref{fig:network} (a), we show a weighted and directed network for the 2022 season. Nodes denote the teams, with the size capturing the win rate. A directed and weighted edge from team $j$ to team $i$ denotes the number of wins of $i$ over $j$ (so that, in network terms, centrality is transferred from the losing to the winning team). To capture season performance by including the quality of the wins, we compute the PageRank centrality for all teams \onkar{CITE}. The PageRank centrality gives a higher score to teams who have themselves beat other teams with a high score. Milwaukee (MIL) and Boston (BOS) concentrate wins against opponents that are themselves hard to beat (redder nodes) occupying positions in the center of a force-directed network layout. On the other hand, Charlotte (CHA) and San Antonio (SAS) are bluer and located near the periphery. In Fig.~\ref{fig:network} (b) shows that PageRank and win rate are highly correlated ($r \approx 0.94$) across all seasons, the expected consequence of schedule balance, with residual scatter capturing the quality adjustment that win rate alone cannot recover.

However, such a network-based approach has certain limitations. For instance, team $A$ can get a higher score by beating a good top-ranked team $B$ in the absence of team $B$'s best player due to injuries in one of the match. Given the frequent nature of injuries in NBA, a simple network-based approach as described above may not reflect the true complexity of the team match-ups. An alternative approach is to dynamically calculate the weights based on the last 10 team outcomes. For example, in the above example, we consider the fraction of matches won in the last 10 games by team $B$ to assign the weight of the link to team $A$. For simplicity, we call this approach Strength-of-Schedule (SoS) adjusted PageRank to reflect the momentum that the teams have carried into the match. In principle, SoS based PageRank should partially overcome the limitation of by reassigning the weights according to opponent quality. However, it introduces its own confounding factor that are particularly difficult to disentangle. For instance, late-season play is structurally biased in two opposing directions. First, eliminated teams may deliberately underperform to improve their draft lottery position, inflating the apparent value of wins against them (a phenomenon known as tanking), while already-seeded contenders routinely rest key players once their playoff position is secured, deflating the cost of losses and the value of wins against them.

Finally, we are interested in predicting the NBA champion and, more generally, playoff performance, based on the various metrics discussed above, ranging from Phil Jackson's rule (operationalized as winning percentage over the first 60 games of a season) to the SoS-adjusted PageRank. We implement an XGBoost classifier algorithm, a gradient-boosted ensemble of decision trees \cite{chen2016xgboost}. Because of the nature of the predicted performance, we need to address a heavy class imbalance. To fix this, we applied balanced sample weights during a cross-validation process. This heavily penalizes mistakes on minority classes (like champions) to reliably evaluate model performance.

In Table~\ref{tab:feature_f1_comparison}, we display the F1-score, a metric used to evaluate the quality of the prediction of the NBA champion with the selected features. Over the available seasons, the Phil Jackson Rule achieves the highest F1-score (0.43), because four out of the five winners met that rule. Conversely, SoS-adjusted PageRank ranks last for champion prediction with an F1-score of just 0.20. Overall, the F1-scores are relatively low because of the difficulty and randomness of the championship route in the NBA. Injuries, unfavourable matchups, and inexperience all play an important role in deciding who will be crowned and are not captured in the features at hand.

In Table~\ref{tab:feature_mae_comparison}, we shift our focus to which feature(s) are the most useful at predicting overall playoff performance. Here we are interested in the mean absolute error (MAE), as we want to favor features that help predict the exact tier of performance the best. For instance, if the model predicted that a team was the champion and in reality they went to the Finals and lost, it did a better job than if it predicted a first-round exit. On this task, PageRank achieves the lowest error with an MAE of 0.99, while the SoS-adjusted PageRank is a close second with an MAE of 1.00. Meanwhile, the Phil Jackson Rule ranks last, yielding the highest error at 1.32.

A comparison of the two tables shows a clear inversion in the usefulness of these indicators depending on the task. The Phil Jackson Rule, is useful alone to identify the champion but is the worst predictor of overall playoff performance, as could be expected from what was shown in Fig. \ref{fig:playoffs}. On the other hand, network metrics struggle to pin down the exact champion but prove to be the best overall predictors of deep playoff runs. All in all, these analyses highlight the strong potential of network-based indicators for assessing team performance in the NBA.

\begin{table}[htbp]
    \centering

    \vspace{0.4em}
    \renewcommand{\arraystretch}{1.15}

    \begin{tabular}{lc}
        \hline
        \textbf{Selected feature(s)} & \textbf{F1-score} \\
        \hline
        Phil Jackson Rule                     & 0.43 \\
        Win Rate                              & 0.37 \\
        Win Rate + Momentum + PJ               & 0.37 \\
        Win Rate + Momentum                    & 0.37 \\
        PageRank                               & 0.35 \\
        Strength-of-Schedule-Adjusted PageRank & 0.20 \\
        \hline

    \end{tabular}

    \vspace{0.1em}

        \caption{F1-Score Ranking of Features to Predict NBA Champion. Higher the value, better the prediction }
    \label{tab:feature_f1_comparison}
\end{table}

\vspace{0.5em}

\begin{table}[htbp]
    \centering

    \vspace{0.4em}
    \renewcommand{\arraystretch}{1.15}

    \begin{tabular}{lc}
        \hline
        \textbf{Selected feature(s)} & \textbf{MAE} \\
        \hline
        PageRank                               & 0.99 \\
        Strength-of-Schedule-Adjusted PageRank & 1.00 \\
        Win Rate + Momentum                    & 1.05 \\
        Win Rate + Momentum + PJ               & 1.06 \\
        Win Rate                               & 1.14 \\
        Phil Jackson Rule                      & 1.32 \\
        \hline

    \end{tabular}
    
        \vspace{0.1em}
        \caption{MAE based Ranking of Features to Predict NBA Playoff performance. Lower MAE denotes a more accurate prediction.}
    \label{tab:feature_mae_comparison}
\end{table}

Note that we have only considered four seasons here to train and test our machine learning algorithm. Given the lack of sufficient data points, expanding the analysis to the full historical record of available NBA seasons is a necessary step before any of these patterns can be taken as conclusive. Ultimately, these patterns remind us that the NBA is not perfectly predictable. Sports are inherently subject to randomness, and the many factors must align for a championship or a deep playoff run to materialize.


\section{In-Game analysis}
\label{sec:game}

The second part of this report focuses on within-game patterns of NBA basketball performance, examining temporal scoring dynamics at the team and player levels, alongside in-game interactions and synergistic player combinations, and their relationship with performance outcomes.

\subsection{Team scoring runs}

\begin{figure}[!h]
    \centering \includegraphics[width=1\linewidth]{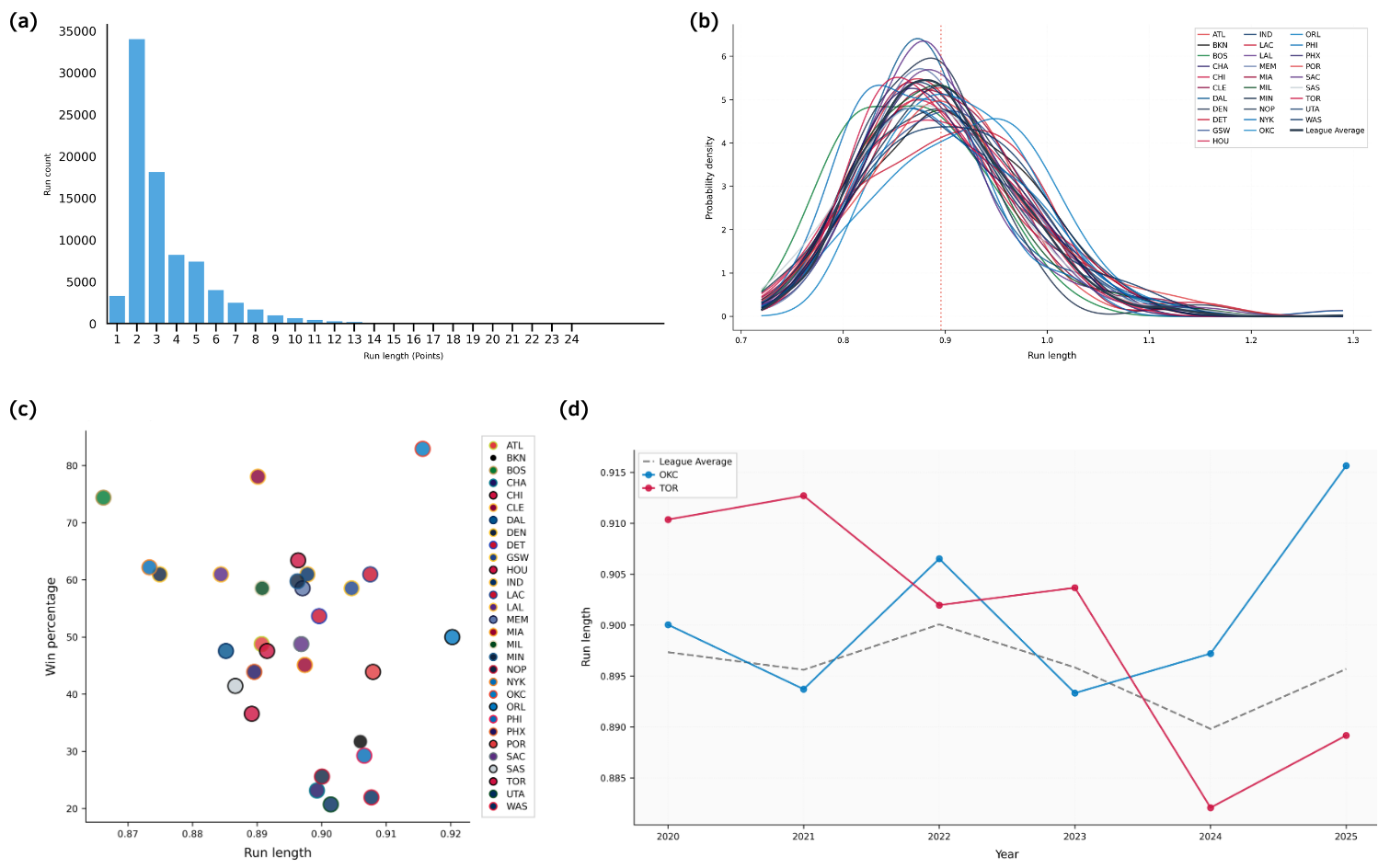}
\caption{
\textbf{Team runs within games}.
\textbf{(a)} League distribution of unanswered run lengths (2025 season). \textbf{(b)} Probability density of team run length over all games across all 30 NBA teams (2025 season). \textbf{(c)} Team wins as a function of team run length.
\textbf{(d)} Evolution of team run length from 2020 to 2025 for the (OKC) and the (TOR) across seasons, illustrating two contrasting temporal trajectories. The dashed line indicates the league-average value in each season.
}
\label{fig:team_runs}
\end{figure}

We start our analysis by looking at individual and team performance within a game. At the team level, we first focus on `unanswered' runs, i.e. consecutive scoring plays without an opponent response. These runs are highlight critical moments in a game where a single team establishes total offensive and defensive control, potentially significantly impacting the outcome of the game. As Fig.~\ref{fig:team_runs}(a) displays, across the 2025 season the length of these unanswered runs varies considerably. The league-wide distribution skews heavily towards short bursts of two to six points, a baseline pattern that reflects the high level of competition found in the league, as well as the inherent randomness that comes with basketball scoring. However, extreme outliers still occur, and this long tail of the distribution is perfectly exemplified by \textit{Dallas’}(DAL) 30 point unanswered run that still resulted in a loss to the \textit{Oklahoma City Thunder} (OKC).

To quantify team-level differences in unanswered run, we introduce (normalised) team run length, defined as average run length of a team, divided by the average run length of a null-baseline obtained by shuffling the order of the scoring events but maintaining the total number of points scored, 

\begin{equation}
    \frac{\langle T_r \rangle}{\langle T_{r_{null}}\rangle}.
\end{equation}

Probability density curves for team run length achieved over all games in 2025 for all 30 teams are shown in Fig.~\ref{fig:team_runs}(b). Values below 1.0 reflect the fact that basketball is a sequential sport where teams take turns on chances to score (while such alternating pattern is ignored in the null-model).
Yet, variability is present across teams in the shape of such curves, with some teams displaying streakier behaviors in their scoring habits. A typical example are the Dallas Mavericks (DAL), whose curve is shifted to the right.

Next, we consider the relationship between team run length and win rate, again for 2025. As shown in Fig.~\ref{fig:team_runs}(c), team run length are not correlated (similar results are obtained for other years). This means that success can be achieved either by teams with high streaky behaviours, and teams with more steady scoring patterns.

Finally, in Fig.~\ref{fig:team_runs}(d) we analyse the evolution of scoring patterns for specific teams over times, compared against a baseline league average. The two teams provide contrasting examples. The run length of the Toronto Raptors (TOR) declined from 2020 ( average run length of $0.911$) following the coaching change, and roster instability through 2024 (run length $0.883$). Conversely, the team run length of the Oklahoma City Thunder increased from 2020 (run length $0.900$), culminating in the 2025 championship associated with run length 0.915.
Further studies are necessary to establish a potential link between team run length, the presence of star players, and winning performances.

\subsection{Lead-change, margins and winning probability}

\begin{figure}[!h]
    \centering
    \includegraphics[width=\linewidth]{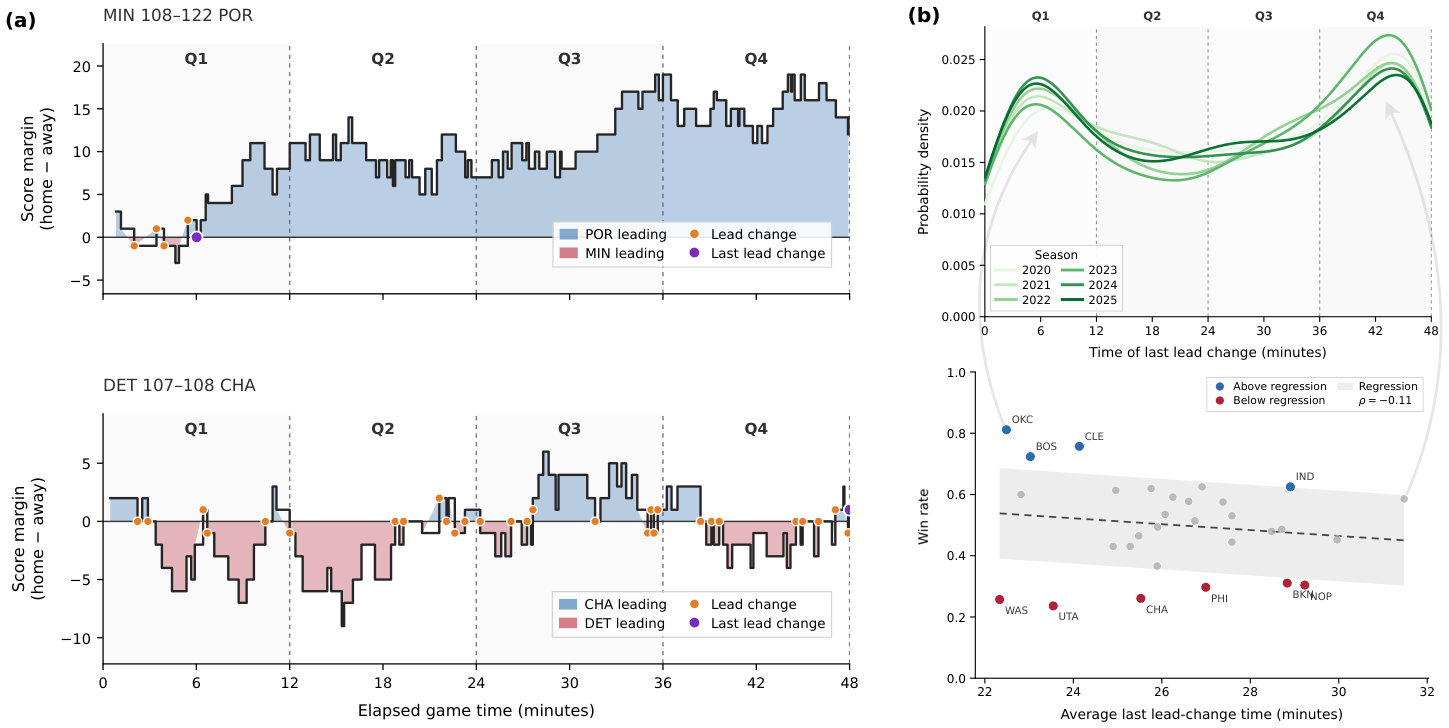}
    \caption{\textbf{Timing and team-level patterns of lead changes within NBA games.}
    \textbf{(a)} Score-margin trajectories for two representative games. Positive and negative values indicate which team is leading, orange markers identify lead changes, and the purple marker indicates the final lead change. The top example illustrates a game (\textit{MIN – POR, 2024}) in which an early final lead change is followed by a sustained advantage for one team, whereas the bottom example shows a more closely contested game (\textit{DET – CHA, 2024}) whose outcome remains unresolved until the final minutes. Vertical dotted lines separate the four quarters.
    \textbf{(b)} Top: distribution of the timing of the final lead change across seasons, expressed as elapsed game time. The distributions show a bimodal pattern, with one peak during the first quarter and a second, more pronounced peak late in the fourth quarter. Bottom: team winning rate as a function of the average timing of the final lead change. The dashed line shows the linear regression and the shaded area its uncertainty. Teams lying far from the regression are more plausibly associated with the early peak, reflecting seasons characterized by relatively less contested games, whereas teams closer to the regression are more plausibly associated with the late-game peak, reflecting seasons in which games remained competitive for longer.}
    \label{fig:within_game_lead_change}

\end{figure}

Next, we examine how game control changes over time. We being by investigating lead changes, defined as the moments at which the team in front is overtaken by its opponent (or the tame become tied). Fig.~\ref{fig:within_game_lead_change}(a) shows two contrasting examples from the 2025 season, both played in November 2024. In the 108--122 loss of the Minnesota Timberwolves (MIN) to the Portland Trail Blazers (POR) only five lead changes occur, all near the beginning of the game. After the final reversal, POR retains the lead and progressively increases its advantage. By contrast, the 107--108 loss of the Detroit Pistons (DET) to the Charlotte Hornets (CHA) contains 30 lead changes, including several during the final minute, describing a closely contested game that remains unresolved until its final possessions. Across the full dataset, games contain on average $6.68 \pm  5.72$ lead changes, 

The number of lead changes describes how frequently control shifts, but not when those shifts end. We therefore define the \textit{last lead change} as the final moment at which the leading team is overtaken. Its distribution across seasons is shown in Fig.~\ref{fig:within_game_lead_change}(b, top). The curves show a similar two-peaked, bimodal-like pattern across seasons (visualized using a Gaussian kernel density estimate) with an early peak in Q1 and a second peak near the end of Q4. The former is associated with games in which one team establishes control early, whereas the latter identifies games that remain open until their closing stages.

We then zoom in on the 2025 season. For each team, we average the timing of the last lead change across its games and compare it with its season winning rate [Fig.~\ref{fig:within_game_lead_change}(b, bottom)]. The regression line summarizes the weak average relationship between the two quantities, while the surrounding interval provides a descriptive reference for distinguishing teams that follow the typical pattern from those lying farther from it. The Oklahoma City Thunder (OKC) and the Boston Celtics (BOS) combine high winning rates with comparatively early last lead changes, suggesting that they often establish control early. The Washington Wizards (WAS) occupy a similar temporal region but have a low winning rate, indicating that an early stabilization of the lead may also reflect an early loss of control. These teams are therefore more consistent with the early component of the distribution, although they lie on opposite sides of the average win-rate relationship. Teams with later average last lead changes are instead more closely associated with the Q4 peak and with games that remain competitive for longer.

\begin{figure}[!h]
    \centering
    \includegraphics[width=\linewidth]{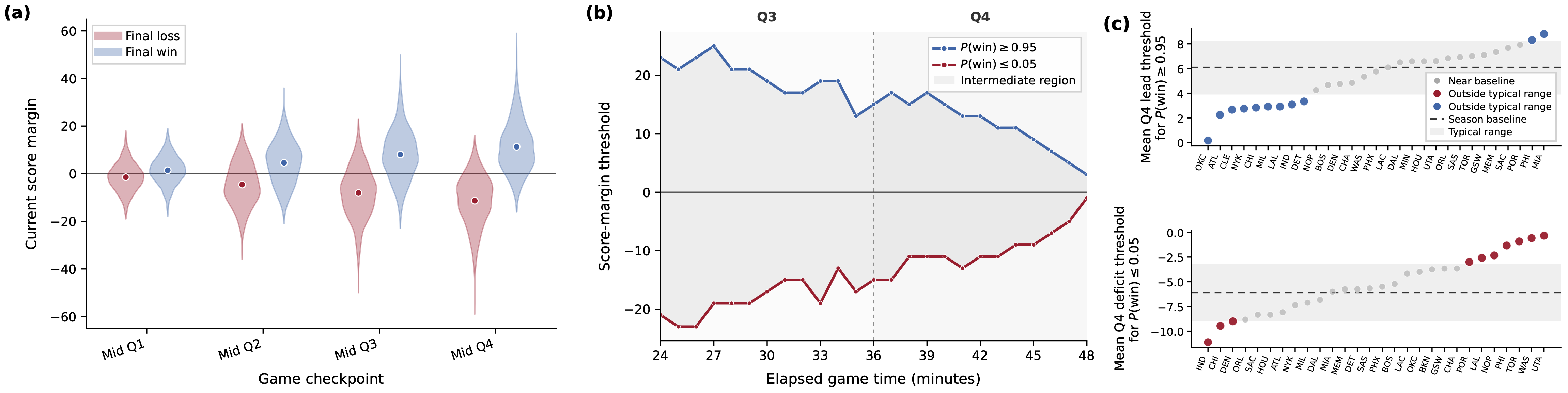}
    \caption{\textbf{Evolution of score margins and outcome-related winning thresholds.}
    \textbf{(a)} Distribution of the score margin at four game checkpoints, separated according to whether the team ultimately wins or loses, in 2025. Points indicate the mean score margin within each distribution, showing how the margins of eventual winners and losers progressively diverge as the game unfolds and the final outcome becomes increasingly reflected in the current score.
    \textbf{(b)}Time-dependent score-margin thresholds associated with a predicted winning probability of at least $0.95$ or at most $0.05$, shown throughout Q3 and Q4. The intermediate region contains score margins for which the game outcome remains less clearly determined in 2025.
    \textbf{(c)} Team-level mean fourth-quarter thresholds 2025 for highly probable wins and losses, compared with the season-level baseline and its typical range. Teams lying outside the typical range are highlighted.}
    \label{fig:within_game_margin_victory}
\end{figure}

The last lead change identifies when leadership stops alternating, but it does not quantify how secure the resulting advantage is. 
We therefore move from this discrete event to the score margin, defined as the difference between the two teams' scores at a given moment. For the 2025 season, Fig.~\ref{fig:within_game_margin_victory}(a) compares the score-margin distributions of eventual winners and losers at four checkpoints, corresponding to the midpoint of each quarter. Early in the game, the two distributions overlap around zero because a small advantage carries limited information about the final outcome. As the game advances, the distribution of eventual winners shifts toward positive margins, while that of eventual losers shifts toward negative margins, showing how the current score becomes increasingly informative about the result.

To quantify this progression, we estimate the empirical winning fraction for each combination of elapsed time and score margin, retaining only combinations supported by a sufficient number of observations. From this relationship, we identify two time-dependent thresholds [Fig.~\ref{fig:within_game_margin_victory}(b)]: the positive margin at which the probability of winning reaches $95\%$, and the negative margin at which it falls to $5\%$. Margins between these thresholds define an intermediate region in which the outcome remains comparatively uncertain. Both thresholds approach zero toward the end of the game because increasingly small advantages or deficits become sufficient to make the final outcome highly probable.

Finally, we zoom in on differences among teams in the 2025 season by averaging these thresholds over Q4 [Fig.~\ref{fig:within_game_margin_victory}(c)]. The season-level regression and its surrounding interval again provide only a descriptive reference, highlighting teams that lie close to or far from the typical league pattern. In the upper panel, a lower positive threshold means that a smaller lead is sufficient to reach a $95\%$ winning probability. OKC is an extreme case in this direction, whereas the Miami Heat (MIA) require a larger-than-typical advantage, suggesting that their moderate leads are less secure. In the lower panel, a more negative threshold means that a team must face a larger deficit before its winning probability falls to $5\%$, indicating a greater capacity to remain competitive while trailing. The Indiana Pacers (IND) provide an example of this behavior. Conversely, thresholds closer to zero indicate that comparatively small deficits already correspond to a highly unfavorable outcome. Together, the two figures describe how often control changes, when it stabilizes, and how the meaning of a lead or deficit depends on both game time and team.

\subsection{Player scoring streaks}

Next, we analyse the patterns of individual player streaks. A player streak within a game is defined as a sequence of consecutive points scored by the same player, possibly interrupted by points from the opposite team, but not by same-team player's points. For instance, if Nikola Jokić scores 12 consecutive points for the Denver Nuggets,  he is on an individual 12 point streak regardless of how many points the opposing team scored during that streak.
Being able to go on scoring streaks is typical of strong offensive players that have the capacity to take over the game for consecutive offensive possessions. 
It is more often observed in crucial moments (e.g. final minutes of close games), when most talented players need to score for their team to win the game or create a significant gap before a break. 

\begin{figure}[!h]
    \centering
    \includegraphics[width=\linewidth]{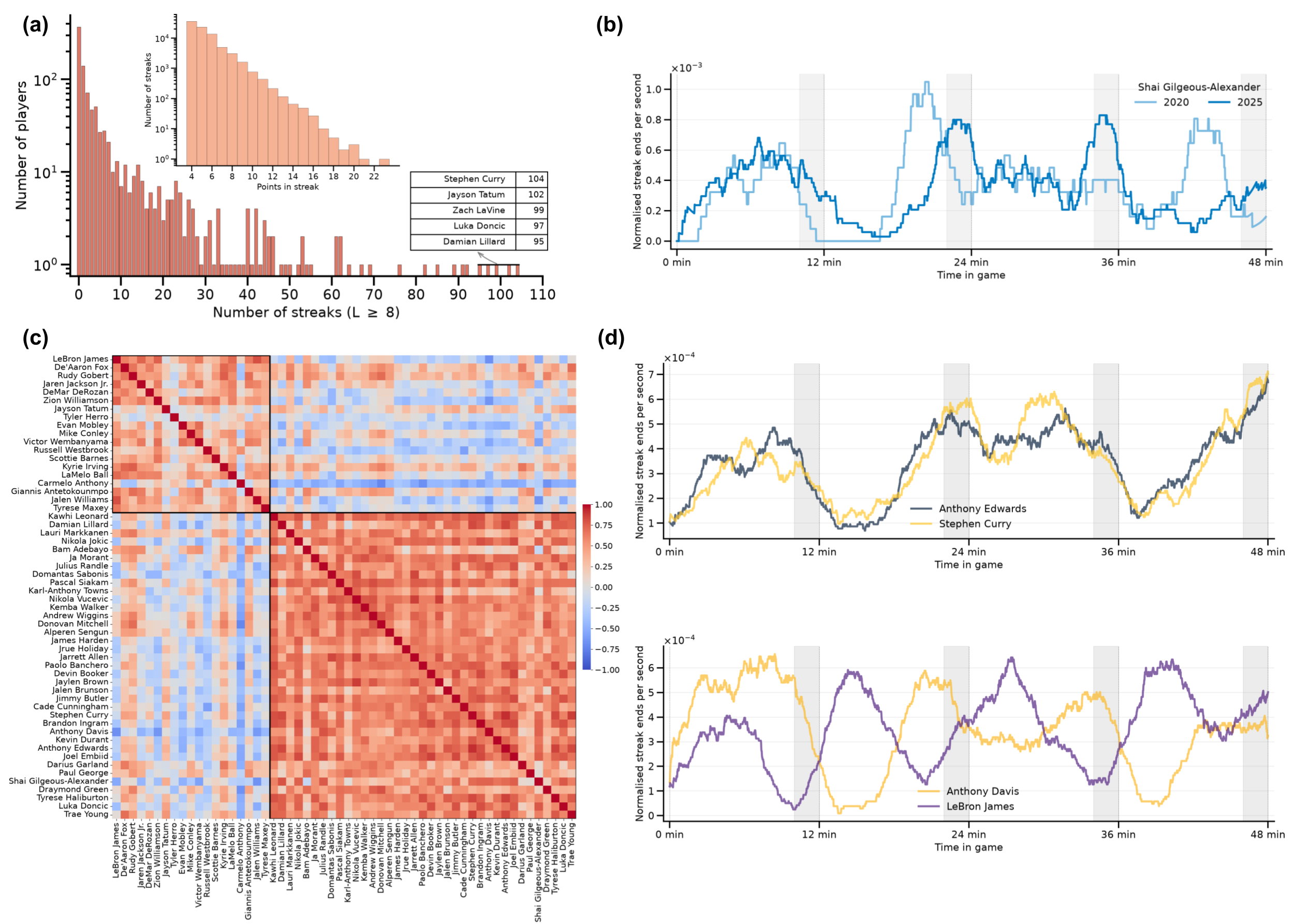}
    \caption{\textbf{Player streak profiles within games.} \textbf{(a)} Distribution of total number of streaks per player, considering only streaks of at least 8 consecutive points, during the 6 seasons considered. Table in the bottom left denotes the top 5 players with the most streaks. Inset: Distribution of streaks by the their point length. 
    \textbf{(b)} Normalised number of streaks per second across a game length for the player \textit{Shai Gilgeous-Alexander} in the seasons 2019/20 and 2024/25 (each curve reports the rolling average with a window of $180s$). Shaded areas correspond to the final $2$ minutes of each quarter. \textbf{(c)} Heatmap of Spearman correlations between each player's streak profile across all the 6 seasons. The set of players chosen corresponds to the ones who were voted as all-star players in the seasons considered. \textbf{(d)} Streak profiles of one of the most correlated pair of players, \textit{Anthony Edwards \& Stephen Curry} $(\rho_S = 0.9172)$, in the top, and one of the most anti-correlated pair of players, \textit{Anthony Davis \& LeBron James} $(\rho_S = -0.6733)$, in the bottom. Shaded areas correspond to the final $2$ minutes of each quarter.}
    \label{fig:within_game_player_runs}
\end{figure}

As depicted in Fig. \ref{fig:within_game_player_runs} (a), a small minority of players ($7.7\%$), has had more than 25 streaks, of at least 8 points, during the seasons considered. Out of those, only the very best players ($2.4\%$) have gone on more than 50 streaks. The table of the top five performers in terms of streaks indeed shows players generally acknowledged as the best players on their respective teams during the seasons at hand. To go further than the raw number of streaks and their length, we can study their occurrence and particularly the moment of the game when each player goes on a streak. A player's streak profile is prone to change from season to season, depending on each player's technical evolution or coaching demands. Fig. \ref{fig:within_game_player_runs} (b) shows the streakiness profile of Shai Gilgeous-Alexander (SGA), in two seasons: 2019-2020 and 2024-2025. 
The displayed patterns are distinct. We note, for instance, that in 2024-2025, SGA had far less streaks in the fourth quarter than in 2019-2020. This is most likely due to the fact that not only he played $1.2$ minutes per game in this quarter, as the average margin of his team that season was $+11.6$ compared to $+1.2$ in 2019-2020, which meant his team did not need him to score imperatively in order to take home the win. However, his scoring habits did not change only in the fourth quarter as a consequence of different usage. Indeed, in 2025 he has also achieved more streaks when in matters the most, before the ends of quarters, to create a large gap with the next team before usually resting at the beginning of the new quarters.  

Finally, in order to identify typical player streakiness profiles, we compare individual players within-game streakiness by computing the Spearman correlation between all pairs of players who were voted at least once as All-Star during the period studied here. Then, we organize the players using hierarchical clustering and identify the two main clusters. As seen in Fig. \ref{fig:within_game_player_runs} (c), each cluster includes players that are highly correlated based on their streakiness profiles, whilst maintaining lower correlations with other cluster's players.
Probing into the typical patterns in each cluster, we identify players in the first cluster as the ones more active in the beginning of each quarter (e.g. LeBron James in Fig. \ref{fig:within_game_player_runs} (d, bottom)), and the ones in the bottom right cluster (that spans the majority of the all-star players) as the ones typically less active in the beginning of the 2nd and 4th quarters (e.g. Anthony Edwards in Fig. \ref{fig:within_game_player_runs} (d, top)), the most common rest time of important players. As an example of these two players with similar profiles, we show in Fig. \ref{fig:within_game_player_runs} (d, top), the streakiness profile of players \textit{Anthony Edwards} and \textit{Stephen Curry}, who were typically not very active in the beginning of the 2nd and 4th quarters, as they usually rest before re-entering back in the game for the final minutes where they do the maximum of their runs. Analogously, we show in Fig. \ref{fig:within_game_player_runs} (d, bottom) \textit{Anthony Davis}, typically more active towards the end of the quarters and \textit{LeBron James}, who concentrates most of its streaks in the early parts of the quarters. Importantly, they have been teammates over the study's time-range and, thus, their performances tend to complement each other.

\subsection{Players' synergy}

In sports contexts, collaborative efforts often exceed the sum of individual contributions, driven by the emergence of synergistic interactions among team members, with collective decision-making, coordination, and role complementarity enhancing overall team performance. Here, we investigate synergistic and anti-synergistic combinations in NBA line-ups. We focus on players who have played in at least 41 games in the season, with an average of at least 20 minutes per game played in all matches where each player of the pair participated into, to avoid drawing conclusions on a very limited sample of playing time.

\begin{figure}[!h]
    \centering
    \includegraphics[width=\linewidth]{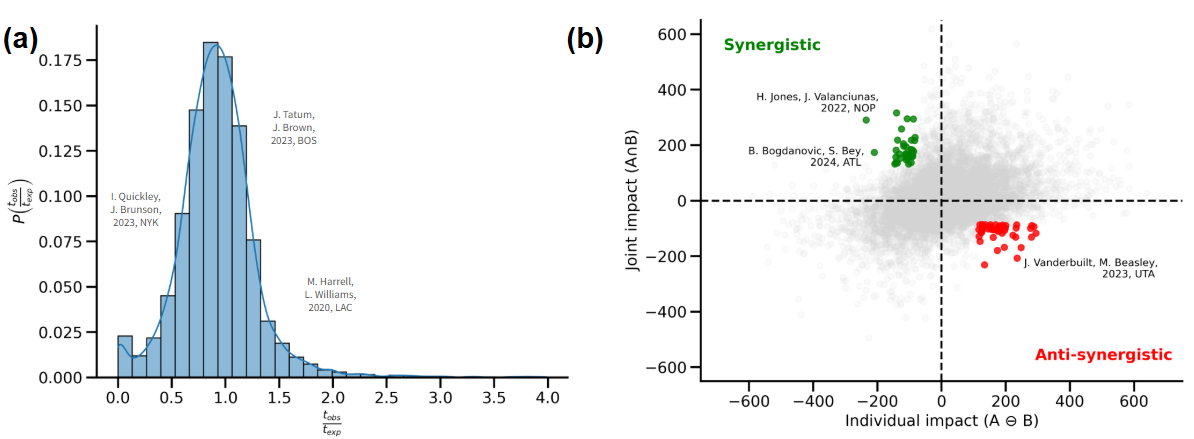}
    \caption{
    \textbf{Synergistic and anti-synergistic interactions between NBA players.}
    \textbf{(a)} Distribution of the ratio between the observed and expected overlap time spent on the court by pairs of players, $t_{\mathrm{obs}}/t_{\mathrm{exp}}$. Values below or above 1 indicate pairs playing together less or more frequently than expected, respectively; selected examples are labelled.
    \textbf{(b)} Joint impact of player pairs as a function of their individual impact when not sharing the court together. Green points highlight synergistic combinations, for which the joint impact is positive despite a negative individual impact, while red points identify anti-synergistic combinations, characterized by positive individual impact but negative joint impact. Grey points represent the remaining player pairs, and selected examples are labelled.}
    \label{fig:within_players_synergy}
\end{figure} 

We first examine the overlap time during which a pair of players are on court simultaneously. For each player pair, we estimate the expected overlap time under the assumption that the two players’ court appearances are independent, and compare this value with the observed overlap time. Fig. \ref{fig:within_players_synergy} (a) illustrates the distribution of observed-to-expected overlap court time ratios $t_{obs}/t_{exp}$ for pairs of players. In general, values below 1 indicate that a pair of players appears together on court less often than expected, suggesting under-utilisation or avoidance of that pairing. In contrast, values above 1 indicate over-representation, meaning that the pair co-occurs more frequently than expected under the null model, consistent with preferential selection or tactical coupling.

By dividing the actual overlap time between two players by their expected joint playing time, a clear trend emerges: most player pairs fall into a baseline category, playing together roughly as much as expected. Often, this is because they are positionally interchangeable: sometimes sharing the floor, sometimes rotating for one another. Most players logging over 20 minutes per game comprise the "starting line-up" and regularly playing usually more than 35 minutes per game. To keep the team consistent these players share the court together a lot, but are also playing with players for the bench to keep the team at a consistent level rather than having a significant drop when all starters are replaced by bench players. In Fig. \ref{fig:within_players_synergy} (a) we highlight Jayson Tatum and Jaylen Brown,  starting duo for the Boston Celtics over the analyzed period: while their observed-to-expected overlap court time ratio is clearly greater than 1, it is also far away from the largest computed values.

Indeed, on the far right end of the spectrum, we typically find bench players who comprise what is tactically known as the "bench unit." Although their individual minute totals are relatively low, meaning a random distribution would dictate they rarely cross paths, their actual co-occurrence is extremely high. Coaches deliberately sub these pairs in together because of their established chemistry and impact.
This dynamic is perfectly illustrated by Montrezl Harrell and Lou Williams during their time with the Los Angeles Clippers. While neither played typical starter minutes (30-40 min), they shared an high proportion of their floor time together as the heart of the bench unit. Conversely, on the left side of the spectrum, we observe players who log significant total playing time but almost never share the court. This pattern typically defines players who occupy the same position as they lack natural tactical complementarity. Then, team success dictates that they be played separately. A clear example of this positional shielding can be seen in pairs of similar profile players like Immanuel Quickley and Jalen Brunson during their time with the New York Knicks in 2023-2024.

On top of simply monitoring shared court time, we investigate the underlying synergy of player duos. The metric we use to quantify this synergy is Net Points, commonly referred to as plus-minus ($+/-$). This metric measures the total points scored by a team while a player is on the court and subtracts the total points scored by the opposing team during those same minutes. It serves as a widely accepted measure of a player's impact on the game's outcome beyond individual box-score statistics. We define a duo as highly synergistic if their individual conditional plus-minus, the net rating they record while playing without the other teammate on the floor, are low or negative, whereas their joint plus-minus spikes significantly when they share the court. This framework highlights players whose game impact is heavily elevated by a specific teammate's presence, often through intangible factors that standard metrics like points, rebounds, or assists fail to capture. Together, such pairs might form an elite defensive duo or an offensive powerhouse that becomes uniquely difficult for opponents to defend. 

Fig. \ref{fig:within_players_synergy} (b) tracks this synergy across all player duos who logged significant minutes over the observed seasons, calling out both the most synergistic and the most anti-synergistic pairings. For instance, Herbert Jones and Jonas Valančiūnas of the 2022-2023 New Orleans Pelicans serve as a prime example of high synergy. Individually, each player carried a negative plus-minus when playing without the other. However, when sharing the floor, they combined for a significantly positive impact. This turnaround is especially remarkable given that the team was pretty average that season (51.2\% win rate), showing a incredibly valuable impact of the duo on their team. 
On the opposite end of the spectrum, we observe instances of negative synergy, such as shown by Jarred Vanderbilt and Malik Beasley during their 2022-2023 season with the Utah Jazz. Individually, both players maintained positive net impacts but, when deployed together, had a distinctly negative impact on the team's performance. 

Combining these co-occurrence and synergy metrics provides a powerful tool for evaluating coaching efficiency. A coaching staff should naturally feature synergistic pairings beyond what their default baseline minutes would dictate. Conversely, visually or statistically dysfunctional pairs should rarely share the floor together, regardless of their individual talents, because their combined chemistry ultimately makes the team worse.

The above analysis can be extended beyond pairs of players, in the spirit of higher-order network analysis~\cite{battiston2020networks}, to observe synergistic patterns beyond dyads~\cite{chowdhary2026team}.

\begin{table}[ht]
\centering
\begin{tabular}{lllll}
\toprule
Players & Season & Team & Plus-minus & $t_{obs}/t_{exp}$ \\
\midrule
Kentavious Caldwell-Pope, Michael Porter Jr., Nikola Jokić & 2023--24 & DEN & 455 & 1.635 \\
Christian Braun, Michael Porter Jr., Nikola Jokić & 2024--25 & DEN & 450 & 1.292 \\
Aaron Gordon, Kentavious Caldwell-Pope, Nikola Jokić & 2023--24 & DEN & 421 & 1.689 \\
Aaron Gordon, Kentavious Caldwell-Pope, Nikola Jokić & 2022--23 & DEN & 415 & 1.769 \\
Jamal Murray, Michael Porter Jr., Nikola Jokić & 2023--24 & DEN & 390 & 1.445 \\
Aaron Gordon, Jamal Murray, Nikola Jokić & 2022--23 & DEN & 389 & 1.536 \\
Isaiah Hartenstein, Jalen Brunson, Josh Hart & 2023--24 & NYK & 376 & 1.004 \\
Bojan Bogdanović, Royce O'Neale, Rudy Gobert & 2020--21 & UTA & 371 & 	1.331 \\
Aaron Gordon, Kentavious Caldwell-Pope, Michael Porter Jr. & 2023--24 & DEN & 357 & 1.723 \\
Jalen Williams, Luguentz Dort, Shai Gilgeous-Alexander & 2024--25 & OKC & 352 & 1.329 \\
\bottomrule
\end{tabular}
\vspace{0.1em}
\caption{Top three-player lineups by plus-minus across seasons from 2019-20 to 2024-25.}
\label{tab:top_triplets}
\end{table}

In addition, Table \ref{tab:top_triplets} reports the highest-performing three-player lineups, ranked by plus-minus, for each season from 2019-20 to 2024-25. Each row represents a unique triplet of players who shared the court in a given season, together with their aggregate on-court performance and usage characteristics. Higher plus-minus values indicate more effective lineups, reflecting a greater margin by which teams outscored opponents during the players’ shared minutes.

The table also reports the ratios $t_{obs}/t_{exp}$ for 10 top-performing triplets across all seasons that we investigated. Similar to the preceding measure, we estimate the expected overlap time under the assumption that the players’ court appearances are mutually independent. The results show that all observed values exceed the baseline, with all triplets playing together more than expected. This suggests that these players not only logged substantial minutes individually, but also spent disproportionately long periods on court together.

Notably, and as expected, all three-player lineups shown here formed core components of highly successful teams in recent years. In the seasons considered, these players were all starters for their respective teams, and their strong collective performance translated into a substantial number of wins. This pattern particularly underscores the exceptional impact of Nikola Jokić during the study period. He appears in all six three-player lineups identified across the seasons analyzed and was voted the league’s Most Valuable Player three times during this period. These results suggest that starting units built around Jokić consistently combined to form outstanding teams.

\section{Discussion}

In recent years, the proliferation of detailed player- and game-tracking data has transformed the landscape of NBA analytics, enabling unprecedented insights into individual performance, decision-making, and team dynamics. By moving beyond traditional box-score statistics, modern analytical approaches leverage large-scale spatiotemporal and event-level data to quantify aspects of the game ranging from shot selection and defensive effectiveness to lineup interactions and tactical strategies. This shift toward data-driven evaluation reflects a broader trend in professional sports, where advanced statistical and computational methods are increasingly integrated with conventional basketball knowledge to inform coaching, player development, and front-office decision-making.

At the season scale, three findings stand out. First,  winning streaks cluster beyond chance for $L \geq 6$, whereas losing streaks do not, suggesting that the ``winning is a habit'' half of Lombardi's dictum has stronger empirical support than its converse. Second, the conventional list of momentum drivers -- home court, rest, recent results -- display higher than expected-by-chance on game outcome, but the celebrated road-trip fatigue only appears past the fifth consecutive away game, while back-to-back schedules emerge as the most consistent depressant of win rate ($44\%$). Third, Phil Jackson's ``40 before 20'' remains a useful champion filter (F1 = 0.43) but a poor predictor of finer playoff outcomes, where a network-based approach such as PageRank centrality (MAE = 0.99) better captures the quality of accumulated wins. A dominant second half of the regular season correlates with deeper playoff runs than an equally dominant first half, although the margin is narrow.

Within games, high variability is present in scoring runs at the team level and over the years. The dynamics of lead-change over time is bimodal, with the last lead change typically occurring either within the first quarter or in the closing minutes of the fourth. Team-specific score-margin thresholds can help quantify how secure a given lead is at each moment of Q3 and Q4. At the player level, long scoring streaks are concentrated in a small elite (7.7\% of players accumulate more than 25 such streaks), and their temporal profiles cluster into two recurrent shapes that match known scoring behavior of superstars. A co-occurrence and plus-minus analysis identifies pairs and triplets whose joint impact substantially exceeds the sum of their individual contributions.

\begin{figure}[!h]
    \centering
    \includegraphics[width=0.8\linewidth]{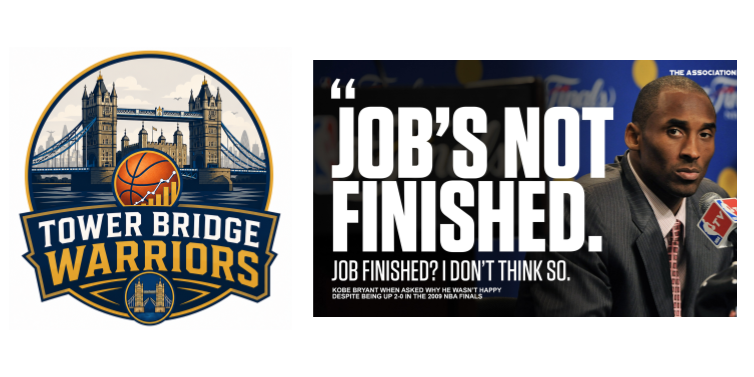}
    \caption{Tower Bridge Warriors logo (generated from ChatGPT) and Kobe Bryant's iconic words after leading 2-0 in 2009 NBA finals.}
    \label{fig:kobe}
\end{figure}

This report was produced in just 72 hours by the Tower Bridge Warriors team as part of the Complexity72h event, held in London in June 2026 (Fig. ~\ref{fig:kobe}). As Kobe Bryant famously remarked during the 2009 NBA Finals, when asked why he appeared unsatisfied despite his team leading the series 2–0, ``Job’s not finished''. In the same spirit, we view this report not as a final destination but as a starting point for further investigation, discovery, and hopefully publication.

\section{Acknowledgements}
This work is the output of the workshop Complexity72h by Complexity Next Gen, held at Northeastern University London, London, UK, 22-26 June 2026. www.complexitynextgen.org/complexity72h/.

\bibliographystyle{unsrtnat}

\end{document}